# Quantum-Well Laser Diodes for Frequency Comb Spectroscopy

MARK DONG[1,2†], MATTHEW W. DAY[2], HERBERT G. WINFUL,[1,2] AND STEVEN T. CUNDIFF[1,2*]

[1]Department of Electrical Engineering and Computer Science, University of Michigan, 1301 Beal Avenue, Ann Arbor, MI 48109-2122
[2]Department of Physics, University of Michigan, 450 Church Street, Ann Arbor, MI 48109-1040
*cundiff@umich.edu

We demonstrate simple optical frequency combs based on semiconductor quantum well laser diodes. The frequency comb spectrum can be tailored by choice of material properties and quantum-well widths, providing spectral flexibility. Finally, we demonstrate the mutual coherence of these devices by using two frequency combs on the same device to generate a radio-frequency dual comb spectrum.

## 1. Introduction

The development of optical frequency comb technology has driven major advances in optical spectroscopy techniques and precision metrology [1]. The high phase coherence between the frequency comb modes facilitates phase coherent connection between the optical and RF frequency domains utilizing techniques such as self-referencing [2]. This highly versatile capability enables many applications such as accurate measurements of atomic transitions [3], molecular spectroscopy [4], optical atomic clocks [5] and arbitrary waveform generation [6]. With the advent of dual-comb spectroscopy [7], multiple frequency combs can be used to improve data acquisition time while maintaining high spectral resolution by eliminating the need for delay lines or diffraction gratings. However, many of the aforementioned applications utilize sensitive and bulky mode-locked lasers for frequency comb sources, including erbium doped fiber [8] and Ti:Sapphire lasers [9]. A truly portable, battery operated chip-scale frequency comb source is a highly attractive platform for greatly expanding the utility of frequency comb spectroscopy beyond the confines of the laboratory [10], with major research efforts in microresonators [11,12] and quantum cascade lasers (QCLs) [13,14]. We have identified and experimentally verified another highly portable and easily manufactured source of frequency combs, which are generated directly from InGaAsP/InP based quantum well (QW) laser diodes. We report several laser diode designs with their measured performance, optical spectra, and RF coherence, demonstrating their great potential to be alternatives to current state of the art frequency comb technology.

## 2. Background

While generating a comb directly from a conventional laser diode has always been an attractive approach, it is not without problems. Attempting to mode-lock the laser diode with a saturable absorbing section to generate short pulses is limited by the intracavity phase distortions due to ultrafast phenomenon originating from the semiconductor carriers [15]. Mode-locking has been more successful with solid state gain media using semiconductor saturable absorbing mirrors (SESAMs) [16] and erbium or ytterbium doped fiber lasers. However, it was previously reported that frequency combs are generated naturally in single-section quantum dot (QD) [17], quantum dash [18], QW laser diodes [19,20], and QCLs [13,14]. Instead of a series of short pulses, these lasers exhibit a quasi-CW output with a time-dependent frequency modulation (FM). The FM solution has several major advantages over short-pulses. First, it avoids the detrimental carrier-induced phase effects inside the laser cavity; second, it is much easier to design and manufacture due to the absence of a saturable absorber; and finally, the output can be phase compensated to transform from an FM signal to a series of short pulses. The versatility of the FM output combined with the simple design of a single laser diode makes these devices highly useful and easily integrated into existing systems.

We have previously theoretically studied the physics of FM comb generation in semiconductor laser diodes [21,22] and found that frequency comb generation should be mostly automatic given proper laser design and material selection. Carrier diffusion should be low to give rise to spatial-hole burning (SHB) and multimode lasing, while intracavity power should be high for coherent, gain mediated four-wave mixing (FWM). It is beneficial to have a large gain bandwidth via inhomogeneous broadening for a broad comb as long as adjacent cavity modes are within the homogeneous gain linewidth so FWM can occur, locking the phases together.

In this paper, we provide an in-depth experimental investigation into the design and performance of these frequency combs. While the previous literature had a focus on telecommunications applications, we pay particular attention to their RF spectra for applications utilizing dual-comb spectroscopy.

### 3. Laser Diode Design

Using our knowledge of the physics of FM combs in semiconductors, we test three InGaAsP/InP QW laser designs for frequency comb generation. The designs are described below, with the semiconductor layer stack detailed in Table 1.

All laser designs are grown on an n-doped InP substrate with 1 μm InP cladding layers and two, 200 nm InGaAsP separate confinement heterostructures (SCH) layers. The first design is very similar to those reported in the literature [19,20] and is our baseline design for evaluating the next two designs. The second design tests our ability to expand the bandwidth by artificially inducing inhomogeneous broadening. By including QWs with different widths whose transition energies are separated by less than the homogeneous half linewidth of the gain to maintain coherence through FWM, we expect the bandwidth of the combs to improve relative to that of a laser with only identical QWs. The third laser design explores the possibility of generating combs at shorter wavelengths than 1550 nm. Despite the fact that spatial-hole burning becomes weaker as the wavelength decreases due to diffusion, comb generation should still be possible at 1310 nm in QW lasers.

The InGaAsP/InP wafers are grown at a commercial foundry and then processed at University of Michigan's Lurie Nanofabrication Facility. The fabrication process uses standard lithography and reactive ion etching to form a 2 μm wide, 900 nm deep ridge waveguide. Before stripping the photoresist, a 600-800 nm layer of silicon oxide is deposited via physical vapor deposition for sidewall passivation and planarization. The wafers are then soaked in photoresist stripper and agitated to facilitate oxide lift-off from the top of the ridges. Once the oxide lift-off is complete, a second lithography step is performed to pattern the metal contacts. We deposit 10 nm / 250 nm of Ti/Au film and perform a second lift-off process. Another layer of Ti/Au is deposited on the bottom side of the substrate to finish the metallization. To complete fabrication, the laser diodes are then hand-cleaved into bars of various lengths (1 mm to 2 mm), forming a Fabry-Perot cavity from the facet reflections. A scanning electron microscope (SEM) image of a completed waveguide is shown in Figure 1, with a finite-element simulation of the ridge waveguide mode shown as an inset. The darker film color is the oxide, while the lighter film on top is the Ti/Au metal contact.

### 4. Single laser diode experimental results

The laser diodes are tested using a probe-station. The hand-cleaved lasers are mounted on a conductive brass sheet on top of a thermoelectric cooler (TEC). Aluminum clamps secure the small laser chip so it has good contact with the brass. A second clamp also secures a thermistor that contacts the chip in order to monitor the temperature. An electric probe on an adjustable stage is brought down on top of the laser to complete the electrical circuit. Laser injection current and the TEC are both controlled by a commercial laser diode and temperature controller. To collect the light from the chip we use an SMF-28 tapered fiber with an AR coating on the tip. We observe a coupling efficiency of 50% - 60%.

Optical spectra from the three laser designs are measured using an optical spectrum analyzer. Figure **2** presents spectra from laser design 1, showing a comb structure spanning several nm, with a repetition rate, or free spectral range (FSR), of 42.36 GHz for the 1 mm laser and a lower FSR of 23.04 GHz for the 1.8 mm laser. The frequency comb bandwidths broaden as the input pump current is increased due to the increasing gain bandwidth, stemming primarily from a larger number of higher energy carriers being injected into the cavity. This interpretation is confirmed by the shape of the spectra broadening toward the higher frequency side, implying gain due to higher energy carriers. However, the measured spectra also exhibit an overall red shift to longer wavelength that is significant (30 - 40 nm from the absorption line center). This shift is likely due to the temperature narrowing of the bandgap [22]. We note that some wavelength tuning of the laser via temperature heating and cooling (through the TEC) is possible, as we observed up to a 5 nm shift by tuning the temperature 5 degrees C.

A comparison of the dual-width QW design 2 to the uniform QW design 1 is shown in Figure **3**. The asymmetric QW optical spectrum shows a bump on the shorter wavelength side that is not present in design 1 lasers, indicating that the higher energy carriers are filling up the narrow, 7 nm QW. We can see that the dual-width QW laser design is superior in terms of bandwidth, particularly on the shorter wavelength side. At an input current of 235 mA, the -10 dB bandwidth is about 1.2 THz. In both the long laser (about 2 mm) and short laser (about 1 mm) the spectrum of the dual-width QW laser is broader. We expect that with better temperature management, we should be able to push the bandwidth even further into the higher frequency side with higher injection current. Moreover, for future experiments, we can cascade more than just two widths of QWs – three or four to expand the total number of frequency comb lines.

For the third design, a typical optical spectrum is shown in Figure **4**. A frequency comb with a bandwidth of about 4 nm, or 0.7 THz, is seen. The spectrum is not as sharp as the ones shown for 1550 nm due to our spectrum analyzer having a more limited resolution at this wavelength. Despite this, we are still able to resolve quite a few comb lines. SHB effects are weaker in the 1.3 μm lasers because of the lower operating wavelength, but we are still able to generate frequency combs fairly consistently over many lasers tested. The comb behavior of the 1.3 μm design is an important experimental confirmation of our theoretical predictions [2] that an optical frequency comb can be generated at this wavelength with QW gain media. While combs at this frequency have been generated with QD materials [17], we have shown that it is also possible with QW lasers.

To characterize the coherence of these combs, we utilize a self-heterodyning technique, which measures the interference of all the comb lines with each other. We use an ultrafast photodiode connected to a high frequency electrical spectrum analyzer to translate the optical spectrum into an RF signal. Typically, a narrow RF linewidth suggests stable comb spacing and low amplitude noise, both indicators of mutual phase-coherence between the optical modes. Measurements of the RF linewidth for design 1 and design 3 are normalized and plotted in Figure 5. The design 1 RF spectra are averaged over 40 acquisitions, while the design 3 RF spectra are not averaged. There is a clear trend in the RF linewidths and signal to noise ratio (SNR): as the input current increases, the SNR greatly increases while the linewidth greatly decreases. This is good for laser operation, as the coherence of the laser improves with increasing laser power (as well as frequency comb bandwidth). The measured RF linewidths are on the order of 100 – 200 kHz, a small fraction of the FSR. The RF signals for design 3 show an interesting behavior: at low injection currents, Figure 5 d), there is not a single RF line but multiple, suggesting that the comb spacing is not constant and varies between the optical modes. As the current injection is increased, the weaker RF lines disappear while the center line narrows and increases in strength. This increase in coherence as power is increased is consistent with the previous RF measurement for the 1550 nm laser. Moreover, the coherence has not been lost in the transition to the asymmetric QW design 2. Figure **6** plots the RF coherence for a 2 mm laser diode of design 2, which maintains the good coherence seen in the designs 1 and 3 lasers.

## 5. Dual Comb Characterization

One particularly interesting aspect of these devices for spectroscopy applications is that multiple devices on the same chip are coherent with each other. In addition, devices on the same chip possess independently tunable spectra, yielding an ultra-compact platform for performing dual-comb spectroscopy in the near-infrared. A diagram of our experimental setup for dual-comb measurements is shown in Figure 7. We use two tapered fibers to collect light from two devices on the same side of the chip. Slight differences in repetition rate and independently adjustable injection currents allow us to tailor the RF dual-comb spectra to be within the measurement bandwidth of the data acquisition hardware. Both these fibers are routed through fiber polarization controllers to maximize the overlap in polarization before a 50:50 beam combiner. Afterwards, the light is sent through a 90:10 beamsplitter, with 10% of the light going to an optical spectrum analyzer (OSA), which allows us to retrieve either individual or combined optical spectra from the frequency combs. Finally, the fiber connects to a photodetector and the signal is acquired at a sampling rate of 2 Gs/s. This configuration allows us to pair any two lasers on the chip as the pitch between devices is such that adjacent devices can be coupled into the tapered fiber.

We test two lasers of design 2 on a single chip. Laser 1 is pumped with ~ 195mA while laser 3 is pumped with ~210mA using separate laser diode current controllers. The individual optical spectra of the two lasers are plotted in Figure 8, showing partial overlap. We are able to measure an RF comb with a frequency spacing of about 9.76 MHz. However, the RF comb does not necessarily begin at 0 Hz due to the carrier envelope offset frequency in each comb, which means that $\nu_{LD1} - \nu_{LD2} = \nu_{RF} = n\Delta\nu_{FSR} + \Delta\nu_{ceo}$. The offset frequency $\Delta\nu_{ceo}$ can be easily adjusted by slightly changing the input current to each laser diode. We modified the offset frequency difference to place the dual-comb signal roughly in the center of our 1GHz detection bandwidth. We acquire 2 MS, over the course of 1ms, giving us an R.F. resolution of 2kHz and a bandwidth of 1 GHz. We observe roughly thirty teeth beating in the R.F. domain, as shown in Figure 8 b.

## 6. Conclusion

We have demonstrated simple, compact, quantum well-based semiconductor lasers whose optical output is a high repetition rate optical frequency comb. The comb is generated by simply powering the device on and allowing SHB and FWM within the devices to generate and phase-lock the comb output. We have presented three different device designs. The first, our most basic, output a comb centered at a wavelength roughly of 1.5 µm. We demonstrated mutual coherence within this comb by performing an R.F. self-heterodyne measurement with a representative comb. Furthermore, we demonstrated the flexibility of this design by incorporating an asymmetric QW gain structure in the second design and changing the center wavelength of the comb devices in the third design. Finally, we demonstrated the potential utility of our devices by generating an R.F. dual comb spectrum which shows roughly 30 distinct comb teeth using two free-running devices on the same chip. In the future, our devices could be employed in compact spectroscopic tools designed for remote sensing or gas concentration sensing.

## 7. Acknowledgements

We would like to thank DARPA SCOUT for funding this research. We would like to thank Prem Kumar for encouraging us to pursue this direction. We also thank the LNF staff for their support in the fabrication of these devices.

*† Current addresss: The MITRE Corporation, 202 Burlington Rd. Bedford, MA 01730. The author's affiliation with The MITRE Corporation is provided for identification purposes only and is not intended to convey or imply MITRE's concurrence with, or support for, the positions, opinions, or viewpoints expressed by the author.*

**Table 1: Construction properties for each design**

| Layer name | | Thickness | Doping | $In_{1-x}Ga_xAs_yP_{1-y}$ |
|---|---|---|---|---|
| n-type InP substrate | | 350 µm | $2\times10^{18}$ cm$^{-3}$ | |
| n-type InP cladding | | 1 µm | $5\times10^{17}$ cm$^{-3}$ | |
| InGaAsP SCH layer | | 200 nm | undoped | X = 0.1712<br>Y = 0.3737 |
| Design 1 | 4x QWs, PL = 1.55 µm | 8 nm | undoped | X = 0.2034<br>Y = 0.7475 |
| Design 2 | 2x QWs, PL =1.55 µm | 8 nm | undoped | X = 0.2034<br>Y = 0.7475 |
| | 2x QWs, PL =1.53 µm | 7 nm | | X = 0.2034<br>Y = 0.7475 |
| Design 3 | 4x QWs, PL =1.3 µm | 8 nm | undoped | X = 0.0808<br>Y = 0.4848 |
| InGaAsP QW barrier layers (5x) | | 15 nm | undoped | |
| InGaAsP SCH layer | | 200 nm | undoped | |
| p-type InP cladding | | 1 µm | $5\times10^{17}$ cm$^{-3}$ | |
| p-type InGaAs contact | | 100 nm | $2\times10^{19}$ cm$^{-3}$ | |

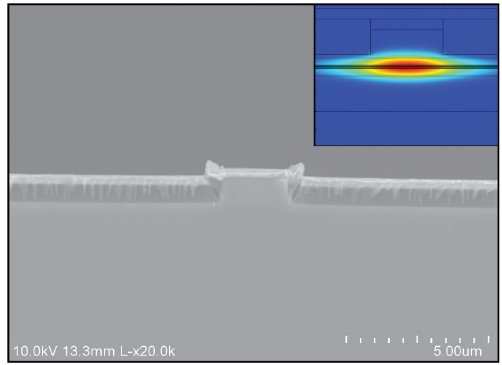

Figure 1: SEM of the ridge waveguide. Inset shows the calculated mode profile

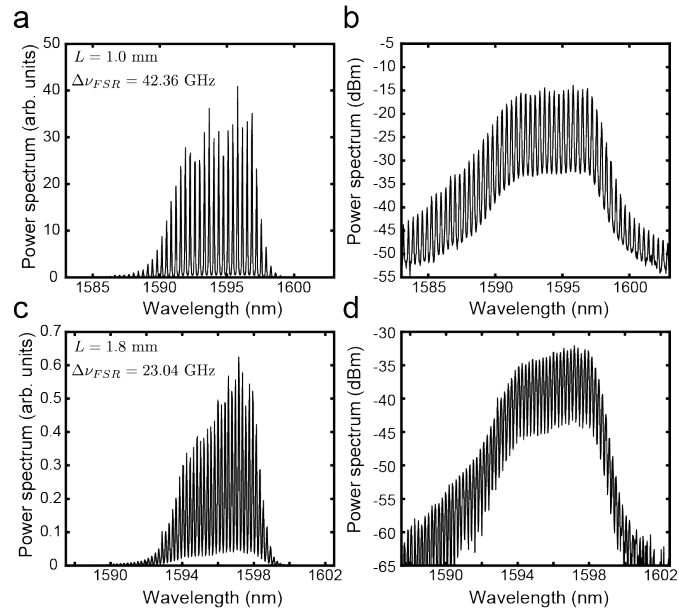

Figure 2: Measured optical spectra of a design 1 laser diode at two different lengths. a) linear and b) log scale optical spectra of a length 1.0 mm laser diode at a pump current of I = 190 mA. c) linear and b) log scale optical spectra of a length 1.8 mm laser diode at a pump current of I = 185 mA.

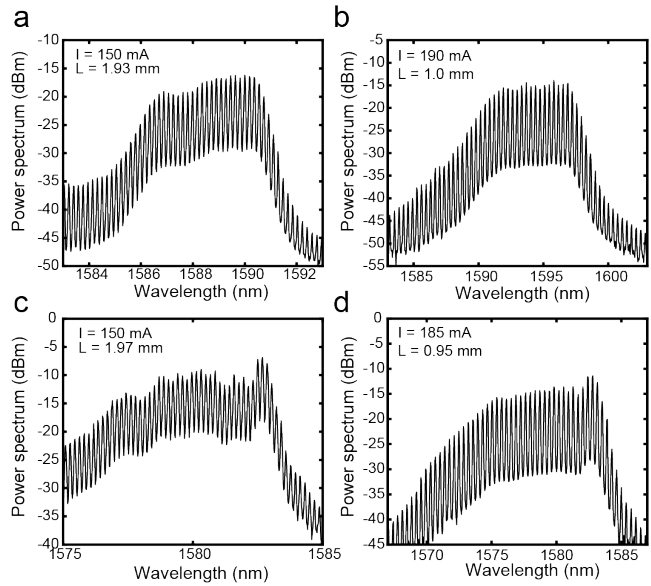

Figure 2: A comparison of the spectra of laser design 1 and design 2. The spectrum of a design 1 laser diode at approximately a) 2 mm and b) 1.0 mm length at similar pump currents. The spectrum of a design 2 laser diode at approximately c) length 2 mm and b) 1 mm at similar pump currents. The dual-width design 2 has a broader spectrum in both case

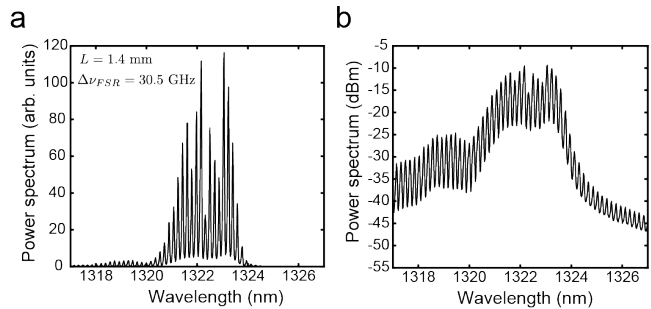

Figure 4: Measured optical spectrum of a design 3 laser diode at pump current I = 135 mA. a) linear and b) log scale plot of the optical spectrum of a length 1.4 mm laser diode, corresponding to an FSR of 30.5 GHz.

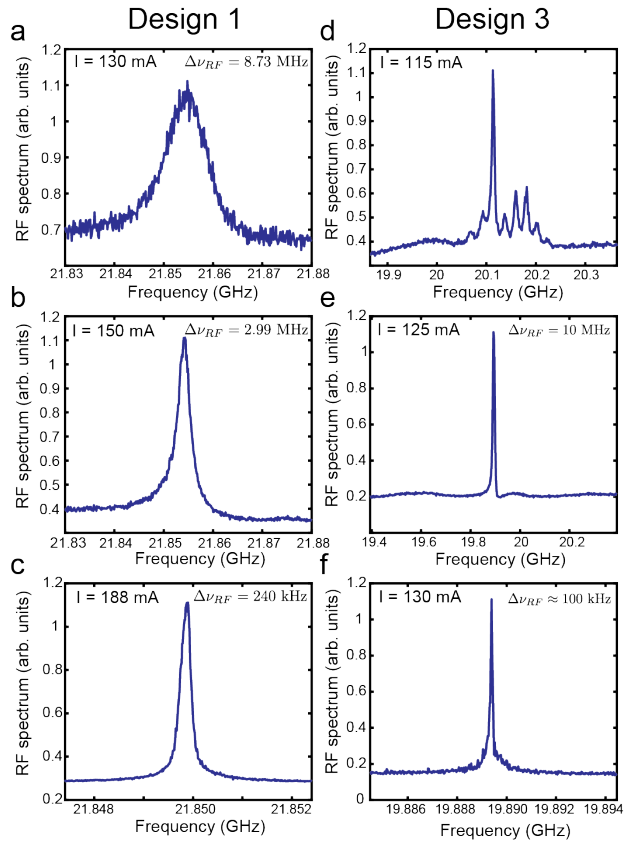

Figure 5: Measured RF spectra of the laser diodes. RF spectra of a design 1, 1.92 mm length laser at a) 130 mA pump b) 150 mA pump and c) at 188 mA pump. RF spectra of a design 3, 2.02 mm length laser at d) 115 mA pump e) 125 mA pump and f) 130 mA pump current.

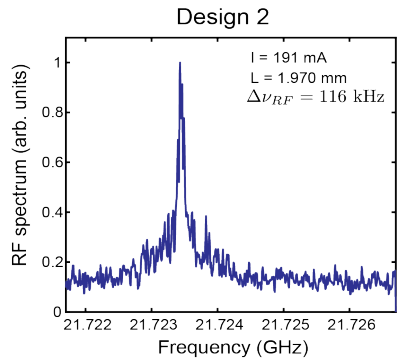

Figure 6: RF spectrum of an approximately 2 mm laser of design 2 at I = 191 mA. The asymmetric QW design RF spectrum maintains the coherence seen in designs 1 and 3.

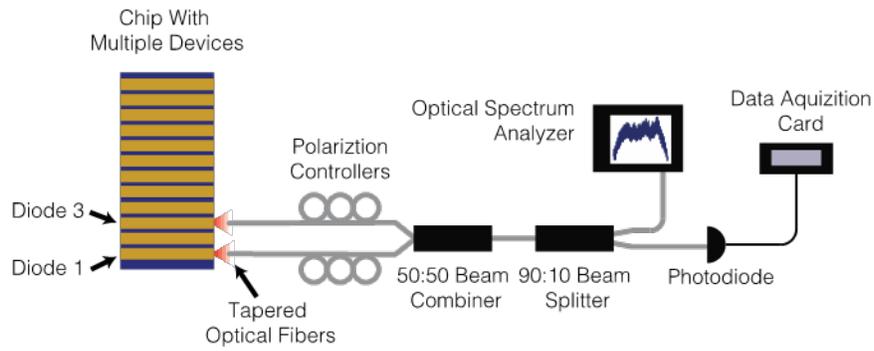

Figure 7: Schematic of the experimental setup for the dual-comb measurements.

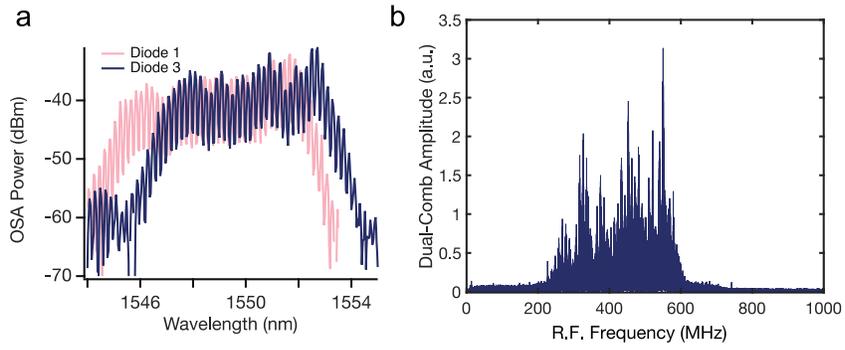

Figure 8: Dual comb spectra: a) an OSA trace of each diode's spectrum, and b) an RF dual-comb spectrum showing roughly 30 separate teeth.